\documentclass[useAMS,usenatbib]{mn2e}

\usepackage{graphicx,amssymb,bm}
\usepackage{subfig,xfrac}
\usepackage{float, Abbrev}
\usepackage[fleqn]{amsmath}  
\usepackage{color}

\usepackage{multirow}

\newcommand{\be}{\begin{equation}}
\newcommand{\ee}{\end{equation}}
\newcommand{\ba}{\begin{eqnarray}}
\newcommand{\ea}{\end{eqnarray}}

\newcommand{\sdm}{\sigma_{\rm DM}}
\newcommand{\sdmi}{\sigma_{\rm  ISO}}
\newcommand{\sdma}{\sigma_{\rm  ANI}}

\def\simlt{\lower.5ex\hbox{$\; \buildrel < \over \sim \;$}}
\newcommand{\cmg}{cm$^2$/g }

\newcommand{\fig}{\begin{figure} \begin{center}}
\newcommand{\efig}{\end{center}\end{figure} }
\newcommand{\figs}{\begin{figure*}\begin{minipage}{180mm} \begin{center}}
\newcommand{\efigs}{\end{center}\end{minipage}\end{figure*} }
\def\simgt{\lower.5ex\hbox{$\; \buildrel > \over \sim \;$}}

\def\dsg{\delta_{\rm DG}}

\usepackage{graphicx} 

\title[Looking for dark matter trails in colliding galaxy clusters]{Looking for dark matter trails in colliding galaxy clusters}
\author[D. Harvey et al]
{David Harvey$^{1}$\thanks{e-mail: {\tt david.harvey@epfl.ch}}, Andrew Robertson$^{2}$, Richard Massey$^{2}$ and Jean-Paul Kneib$^{1,3}$ \\
$^{1}$Laboratoire d'Astrophysique, EPFL, Observatoire de Sauverny, 1290 Versoix, Switzerland \\
$^{2}$Institute for Computational Cosmology, Durham University, South Road, Durham, DH1 3LE, UK\\
$^{3}$Aix Marseille Université, CNRS, LAM (Laboratoire d'Astrophysique de Marseille) UMR 7326, 13388, Marseille, France}

\begin{document}

\date{Accepted ---. Received ---; in original form \today.}

\pagerange{\pageref{firstpage}--\pageref{lastpage}} \pubyear{2013}

\maketitle

\label{firstpage}

\begin{abstract}
\noindent If dark matter interacts, even weakly, via non-gravitational forces, simulations predict that it will be preferentially scattered towards the trailing edge of the halo during collisions between galaxy clusters. 
This will temporarily create a non-symmetric mass profile, with a trailing over-density along the direction of motion.
To test this hypothesis, we fit (and subtract) symmetric halos to the weak gravitational data of 72 merging galaxy clusters observed with the {\em Hubble Space Telescope}.
We convert the shear directly into excess $\kappa$ and project in to a one dimensional profile.
We generate numerical simulations and find that the one dimensional profile is well described with simple Gaussian approximations.
We detect the weak lensing signal of trailing gas at a $4\sigma$ confidence, finding a mean gas fraction of $M_{\rm gas} / M_{ \rm dm} = 0.13\pm0.035$.
We find no evidence for scattered dark matter particles with a estimated scattering fraction of $f=0.03\pm0.05$. 
Finally we find that if we can reduce the statistical error on the positional estimate of a single dark matter halo to $<2.5\arcsec$, then we will be able to detect a scattering fraction of $10\%$ at the $3\sigma$ level with current surveys.
This potentially interesting new method can provide an important independent test for other complimentary studies of the self-interaction cross-section of dark matter.

\end{abstract}


\section{Introduction}
The current best-fit model of the Universe assumes that 84\% of all the matter is in the form of some unknown, non-baryonic `dark matter' (DM) \citep{planckpars}.
In the Standard Model, DM is assumed to be a weakly interacting massive particle (WIMP) that acts collisionlessly.
Despite the relatively simplistic assumptions, cosmological simulations of cold dark matter (CDM) have been able to reproduce the large scale structure of the Universe up to 10\% at a $k=1h$/Mpc \citep{EvolutionLSS,BOSS_clustering,2dgf,vipers}. 
However, conclusive observational evidence of a particle DM is yet to be confirmed \citep[e.g.][]{LUX,xenon100,fermidetect,LHCconstraints}.

Although broadly successful, simulations of collisionless CDM continue to predict many more large galactic sub-halos that should form stars \citep{toobigtofail}, and cusps in dwarf galaxies that appear to harbour cores \citep{corecusp}. Moreover, discrepancies have also been seen in clusters where the gradient of the inner density profile departs from the expected NFW \citep{NFW, densityProf3}.
Such inconsistencies have been attributed to insufficient complexity when simulating astrophysical feedback processes such as supernova and active galactic nuclei \citep{densityprofSIDM}.
However, previously proposed extensions to the Standard Model of particle physics would also resolve these discrepancies.
For example, cusps would be removed if dark matter matter were lighter allowing dark matter particles to free-stream out of potentials \cite[e.g.][]{WDM,WDM2}.
Alternatively a non-zero self-interaction cross-section can cause the formation of a core, reducing the central densities of galaxies and removing cusps \citep{SIDMSim,SIDMSimA,SIDMSimB}.

\subsection{ Constraining $\sdm$ using colliding galaxy clusters}

The only way to constrain the {\it self}-interaction cross-section of dark matter (SIDM) is with astronomical observations where dark matter is present in sufficient quantity to be detected gravitationally.
Several methods have been used to constrain different models of self-interacting dark matter at collision velocities of $\sim1000$km/s. The steady-state shape and sphericity of relaxed clusters nominally yields tight constraints \citep{SIDMtest,SIDMSim}, but is subject to degeneracy between the distribution of dark matter and baryons \citep{densityProf2,densityprofSIDM}.
Galaxy and galaxy cluster mergers have yielded what has become considered the most robust constraints on the cross-section of dark matter.

The first constraints derived from colliding galaxy clusters were placed by measuring the displacement between hot X-ray emitting gas and dark matter in the Bullet Cluster \citep{bulletcluster} where they concluded that $\sdm<1.25$\cmg. Subsequent studies used the same assumption and method finding constraints of $\sdm<3$\cmg \citep{A2744}, $\sdm<4$\cmg \citep{minibullet} and $\sdm<7$\cmg \citep{musket}, however they were limited by sample size and the unknown state of a single merger. 
\citet[][hereafter R08]{impactpars} simulated the Bullet Cluster collision using elastic collisions with an isotropic scattering angle. This resulted in the then tightest constraints of $\sdmi<0.7$\cmg.

Most recently a study of colliding galaxy clusters attempted to circumvent the unknowns in cluster mergers by creating a sample and averaging offsets over many different scenarios \citep[][hereafter H15]{Harvey15}.
One advantage of exploiting the positional estimates of halos to study the cross-section is that the positional estimate of halos via weak gravitational lensing is not affected by many of the problems that are inherent in weak lensing \citep{HARVEY13,Harvey14}. 
The study placed constraints on the long range interaction of $\sdma<0.47$\cmg at the 95\% confidence limit.

In this letter we develop a novel method to observe the potential particle interactions  of dark matter that is independent of the particle physics.
We will outline the proposed new method, test it 
and then apply our method to data and present our results.

\section{A new method to constrain the cross-section of dark matter}



Recent simulations have found that an isotropic scattering in the dark sector can produce a secondary population of particles trailing their parent halo during the collision of two galaxy clusters \citep{SIDMModel,darkgiants}.
The resulting asymmetry in the density profile of dark matter would not be accounted for by any symmetric, parametric model attempting to fit the data. 
As a result, once the best-fitting model has been subtracted off, a residual density correlated with the axis of motion will exist and hence observable if stacked over many merging events.
Furthermore, the mean residual density perpendicular to this axis should be zero since although in individual cases the fit will not describe exactly the distribution of dark matter, there is no known physical process that can induce correlated excess density in the perpendicular direction.
Throughout this paper we will refer to the axis parallel with axis of collision, $r_{||}$ and the axis perpendicular to the axis of collision, $r_\times$, where the axis of collision is the vector joining the dark matter to the gas.
%
Specifically the process is as follows, we also show the method diagrammatically in Figure \ref{fig:draw_demo}.

\fig
       \includegraphics[width=0.45\textwidth]{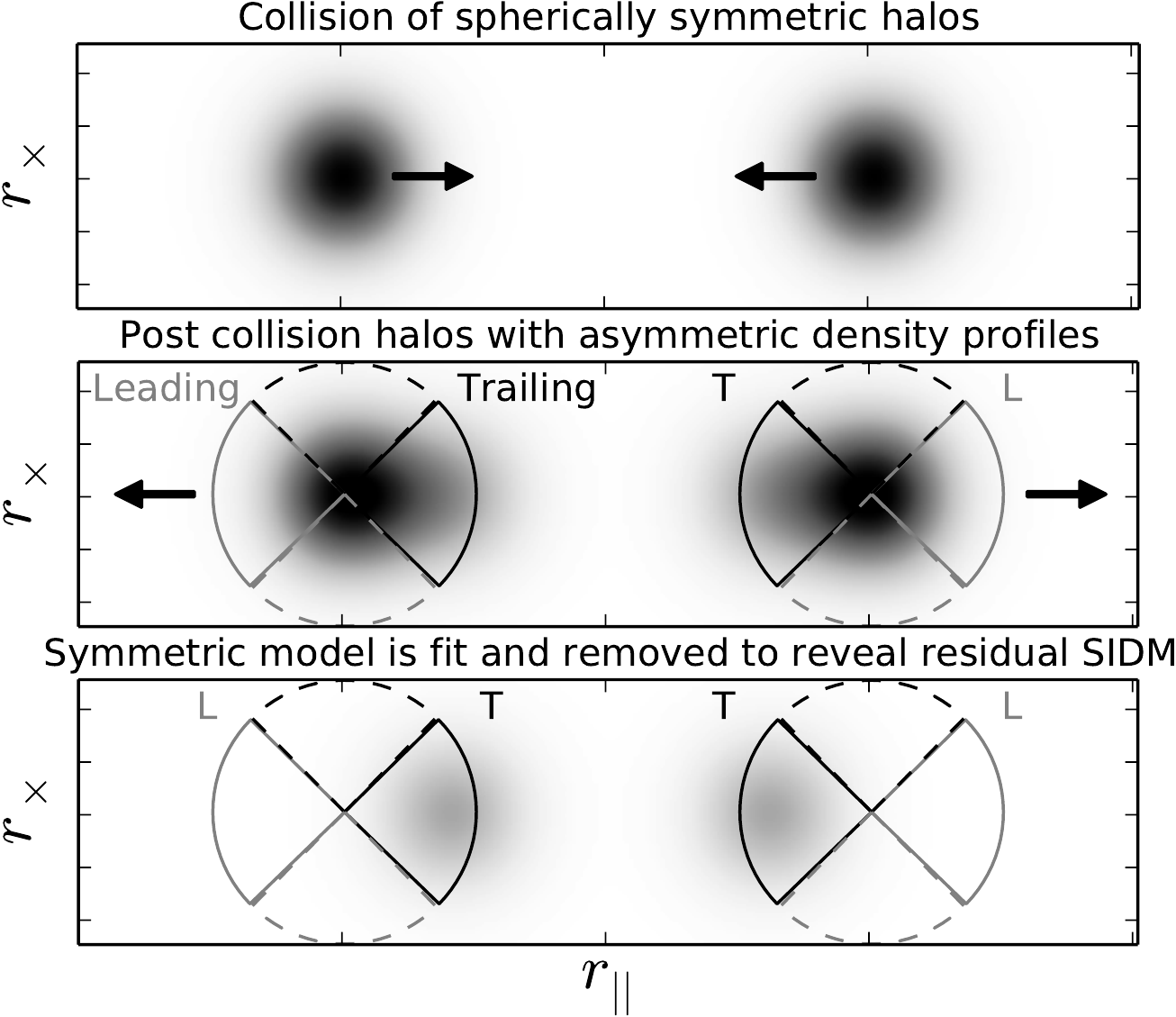} 
       	\caption{ During a collision of two galaxy clusters (panel 1), self-interacting dark matter will be scattered preferentially towards the trailing edge of the halo causing an asymmetry in the profile of dark matter density (panel 2). 
	Using weak lensing, a symmetric, parametric model can be fit to the dark matter density profile and removed to leave any residual dark matter not accounted for by the fit (panel 3). 
	By rotating all the mergers into the same axis of collision, (defined by the dark matter to gas vector) the signal can be stacked over many mergers and extract any potential evidence for interacting dark matter.
	The solid cones show the regions we bin along the $r_{||}$ axis and the dotted cones the regions we bin along the $r_{\times}$, and the definitions of the leading and trailing edge.}
	\label{fig:draw_demo}
\efig
 \subsection{Extracting the asymmetric component}\label{sec:modelsub}
 During the collision of two galaxy clusters (panel 1 of Figure \ref{fig:draw_demo}), dark matter interactions will cause particles to scatter towards the rear of the halo producing a secondary population of dark matter (panel 2 of Figure \ref{fig:draw_demo}).
 
To extract this second population, first we identify the number of large scale ($\sim$Mpc) dark matter halos within the cluster merger. 
We define a merging cluster by identifying X-ray clusters with bimodal emission, and then determine the number of dark matter halos by the number of resolved X-ray emitting gas halos from the Chandra X-ray Observatory data \cite[see][]{Harvey15}. 
This means we assume that any bound halo that is in a state of merger will still retain its gas halo.

To estimate the large scale dark matter distribution we use weak gravitational lensing.
For reviews please see \citet[e.g.][]{BS01,RefregierRev,gravitational_lensing,MKRev,HoekstraRev}.
We first measure the shapes of distant galaxies that have had their isophotes altered by the distribution of matter in the cluster along the photons geodesics.
With theses shapes and using analytical descriptions of dark matter halos we use an open-source program called {\sc Lenstool} to fit a dark matter model to the data.
Since these fields have colliding galaxy clusters  with multiple components, we simultaneously fit multiple, elliptical NFWs \textbf{(one for each large scale halo identified in the X-ray emission)} \citep{NFW}, which are symmetric along the major and minor axes. The density profile of an NFW is given by
\be
\rho/\rho_0 = [ x (1 + x)^2]^{-1},
\ee
where $x=r/r_{\rm s}$, the radial distance normalised to the scale radius of the cluster, which itself is related to the virial concentration, $c_{\rm vir}$, and virial radius, $r_s = r_{\rm vir}/c_{\rm vir}$.
Each dark matter halo fit therefore has six free parameters: position (right ascension, declination), virial mass, NFW virial concentration, ellipticity and position angle. 
Since we are only fitting the large scale halo, and not galaxy scale haloes, we do not assume that light traces mass, only that the main component of the halo follows an NFW.
Moreover, we do not fix the mass-concentration relation as this will most likely not apply in the case of merging halos. 
Like all mass mapping, the derived lensing model can be subject to mass-sheet degeneracies \citep{mass_sheet_1,mass_sheet_2}. 
This will need to be considered in any future interpretation of scattered dark matter, but is not currently an issue while we are simply looking for a detection.

Using the best fitting parametric mass model for the merging cluster, we remove this signal from the data, producing a residual map.
We do this by projecting the source galaxies that are in the `image plane', back to the `source plane', effectively de-lensing the effect of the cluster. 
Hence we remove the signal directly from  the shear and not the mass density.
If the fit is a good one, the `source plane` galaxies now should have no residual gravitational shear signal (and should be completely randomly orientated).
From these `source plane' galaxies we can generate a residual map of the dark matter that is not accounted for by the fitted model (panel 3 of Figure \ref{fig:draw_demo}).

   \figs
                     \includegraphics[width=\textwidth]{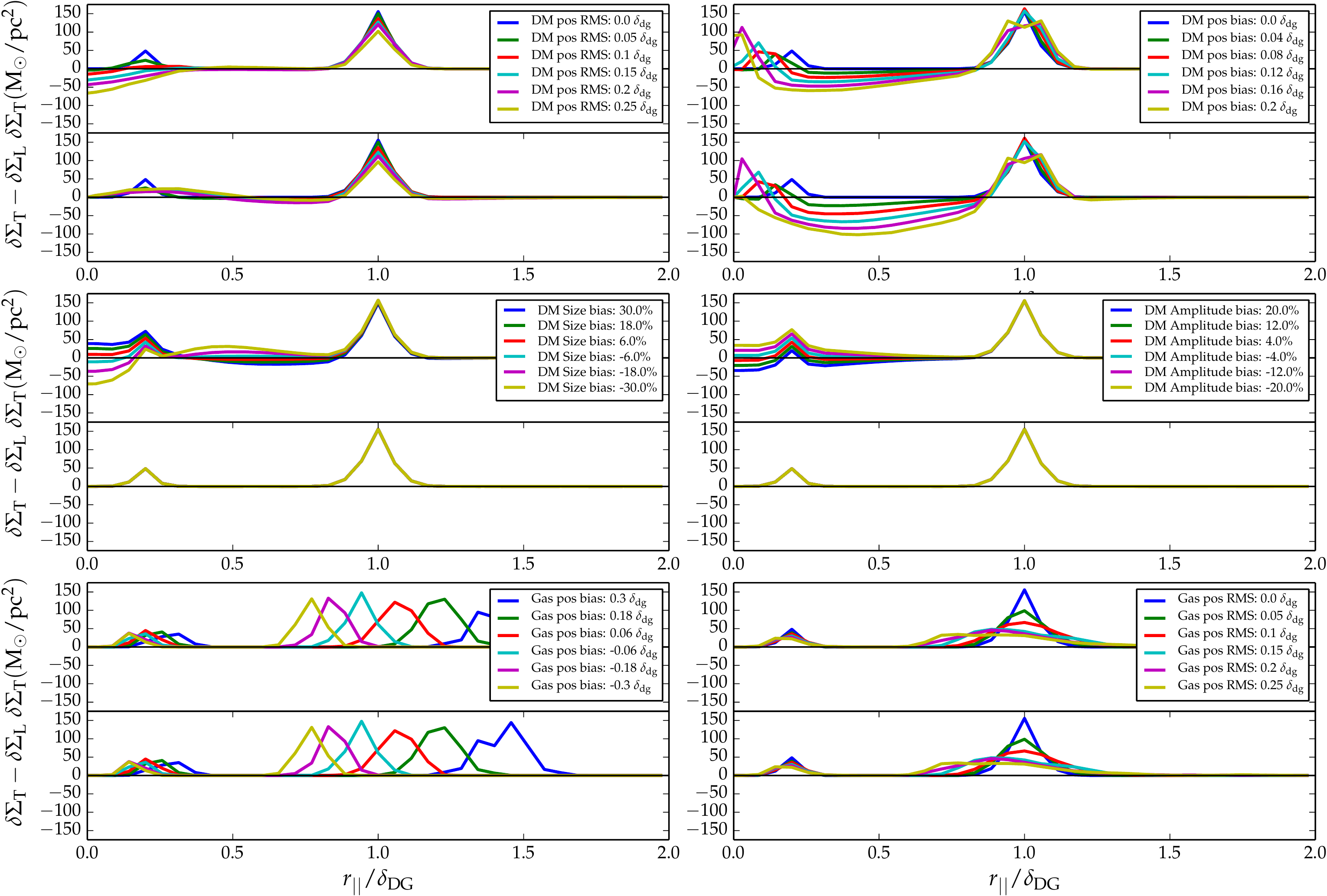} 
       	\caption{ We numerically simulate the method with a 2-dimensional density field test-bed, with a large-scale dark matter halo, a gas halo and a second population of scattered particles. We vary different sources of uncertainty and study how this may affect its potential detection. 
	Top left panel shows the affect of different statistical uncertainties on the estimated peak position of the dark matter halo (in arc-seconds). 
	The middle left panel shows the affect of a systematic bias in the estimate of the size/characteristic scale radius of the dark matter (in fractional error).
	The bottom left panel shows the effect of a of mis-centering the gas halo (in arc-seconds)
	The top right panel shows the effect of a bias in the position of the dark matter halo (in units of $\dsg$)
	The middle right panel shows the affect of systematically under or overestimating the normalisation / amplitude of the dark matter halo (in fractional error) .
	The bottom right panel shows the affect of statistical error in the gas halo position (in arc-seconds).}
	\label{fig:bias_constraints}
\efigs

\subsection{Stacking the signal}\label{sec:modelstack}
In order to detect the potentially very small excess in mass caused by asymmetry in the cluster profile, we stack the galaxies from many fields of clusters.
Given that each cluster has a different merging velocity and direction, first we define a frame of reference for the halo, where the origin is the best fit peak position of the dark matter halo, and the direction of the x-axis is the vector between this origin and the position of the X-ray emitting gas. We define this axis as the axis of collision, $r_{||}$ and its orthogonal axis, $r_{\times}$.
We rotate all the galaxy positions into this coordinate frame and then normalise all the distances from the origin to the magnitude of the vector between the dark matter peak position and the X-ray emitting gas, $\dsg$. This means that a $r_{||}/\dsg=1$ is the separation between gas and dark matter.

By normalising to $\dsg$, we will mitigate any inherent uncertainties associated with the collision impact parameter. Firstly because it will down weight those interacting halos that have not separated their halo, and secondly, any small shift in asymmetry will be fractionally the same whether the halo has gone through a direct core-core passage, or whether the cluster collision was a minor deflection \citep{Harvey14}. This should mean that any second population of particles will scatter to the same point along the vector between the dark matter and the gas. 

\subsection{Creating a one-dimensional density profile}\label{sec:modelproj}
Having stacked many galaxies from an ensemble of clusters into the same reference frame, we create a surface density map using the Kaiser-Squires formalism \citep{KS93}, which relates the observed shapes of galaxies to the projected surface density along the lines of sight. This gives us a two dimensional map in the reference frame of the collision axis, again normalised to the magnitude of the dark matter-gas separation.  Since this requires a regular grid, we create a 2-dimensional density map with a bin width of $\delta_x=\delta_y=0.1\dsg$, we also Gaussian smooth the map by $0.3\dsg$. We verify that the Gaussian kernel has no impact on the results, only that it smooths out some of the noise due to discrete pixels in the map.

Finally, we project the two dimensional free-form residual surface density map into one dimension along the axis of collision, examining the profile in radial bins along the $r_{||}$ axis, taking care only to bin up to the $45^\circ$ axis dividing $r_{||}$ and $r_\times$.
By normalising each cluster by the distance between the dark matter and gas, it is unclear exactly how the errors will propagate through to the final result. To quantify this we create simulations of the method and pass them through the analysis pipeline to the see the effect.

\section{Systematics and error propagation}\label{sec:errors}
This method relies heavily on accurate and precise models of the large-scale dark matter halo of a colliding galaxy cluster.
However, the best fitting models derived from weak gravitational lensing data are sensitive to a variety of statistical and systematic errors. 
In order to estimate and understand how these uncertainties propagate through to our final result we conduct a number of numerical tests.


Modeling the process to understand the key systematics is very difficult. The process of fitting the shear with NFW profiles and then reconstructing the residual map using a free form mass mapping is very time consuming and processor intensive. In order to efficiently explore the parameter space for possible biases we model this process with 2D Gaussian halos for the projected surface density of our halos. We will find later than the data well fits a Gaussian and does not prefer a more complicated model.
Hence we construct a test-bed consisting of a two dimensional density field with two Gaussian halos: one located at the centre of a field 3'x3' mimicking a dark matter halo, and a second halo $40\arcsec$ to the west of the halo mimicking a gas halo, similar to that of the Bullet Cluster.


The gas halo is scaled by  the cosmological baryon fraction ( $\Omega_{\rm B}/\Omega_{\rm DM}=0.17$ ) to the dark matter halo \citep{planckpars}. We then add in a third Gaussian halo into the simulation, mimicking a second population of scattered particles. In \cite{SIDM_bullet}, they find that during a collision about $\sim23\%$ of particles scatter for a cross-section of $1$cm$^2$/g within $400$kpc. This scattered fraction is for particles that initially belong to the smaller (bullet) halo of the Bullet Cluster, that scatter from a particle belonging to the main halo during the collision. The value is sensitive to different halo masses and concentrations, but we take this as representative of large galaxy clusters. Assuming that the number of particles which scatter increases linearly with cross-section, we directly compare this method to the constraints gained in \cite{Harvey15} of $\sdm<0.47$\cmg and hence we create a second Gaussian, $10\%$ of the main halo, situated $8\arcsec$ from the main halo, (or $0.2\dsg$). 
We then vary different sources of uncertainty and identify how each affects the detection of the second population of particles.

In order to replicate the model fitting and subtraction procedure (see \ref{sec:modelsub}),  we subtract a different Gaussian from the simulated density field and analyze the data exactly as we have outlined in the previous section. 
We then introduce various uncertainties into the Gaussian model that we use to subtract off the data and see how this affects the results.
These uncertainties include, random (statistical) and systematic error in the estimate of the peak position of the dark matter halo, a systematic bias in the estimate of the size of the dark matter halo and the amplitude of the dark matter halo and a statistical and systematic bias in the estimated position of the gas halo.
We then free these different parameters up, run 100 Monte Carlo realisations and stack each realisation (see \ref{sec:modelstack}).
This simulation will address how uncertainties in the position of each halo, propagates in to the uncertainty on the normalisation length, $\dsg$, and ultimately the results.

Having stacked each Monte Carlo realisation, we then project the 2D residual density distribution into a one dimensional profile along the estimated axis of collision $r_{||}$ (see \ref{sec:modelproj}). 
  \figs
                     \includegraphics[width=0.49\textwidth]{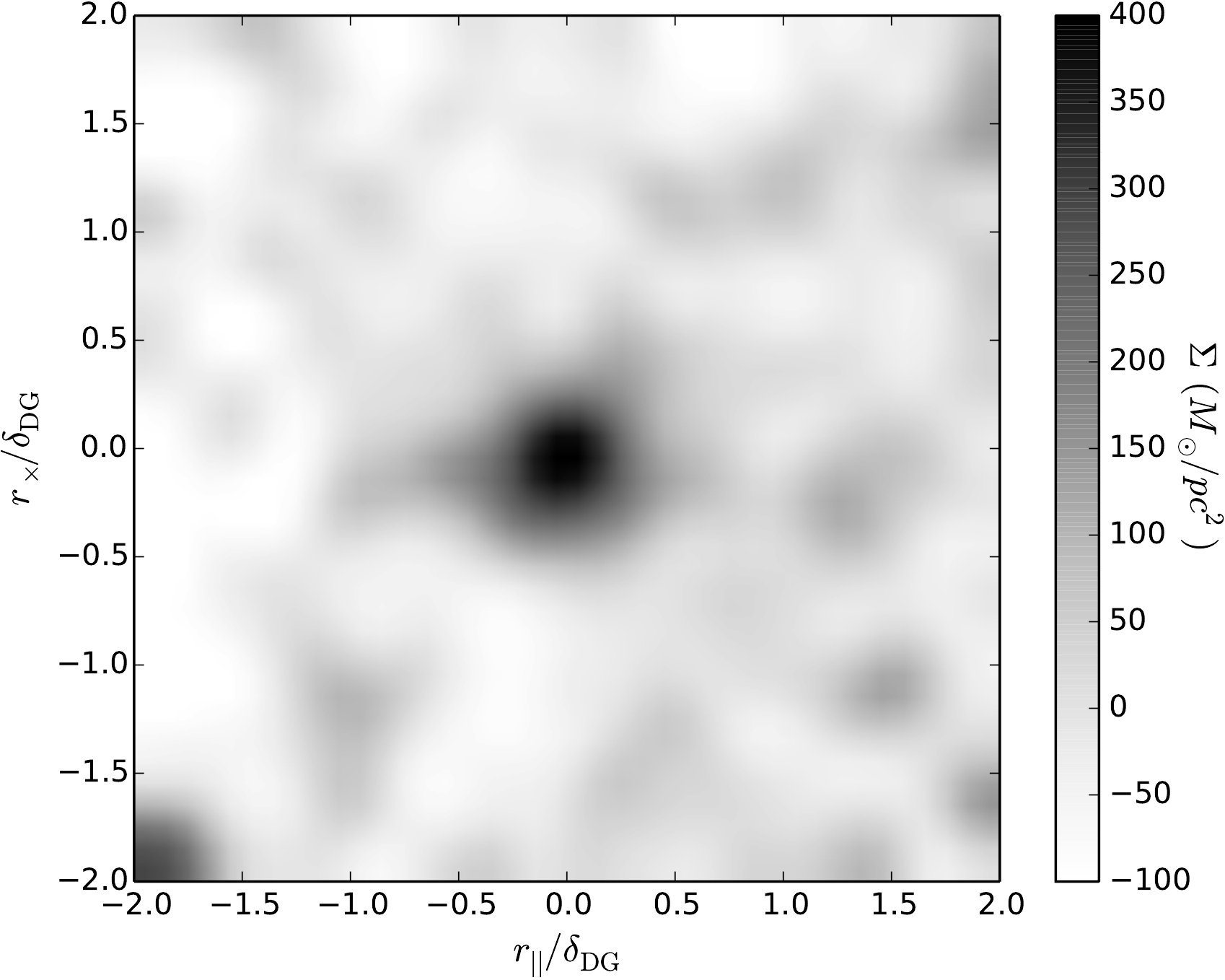} 
       \includegraphics[width=0.49\textwidth]{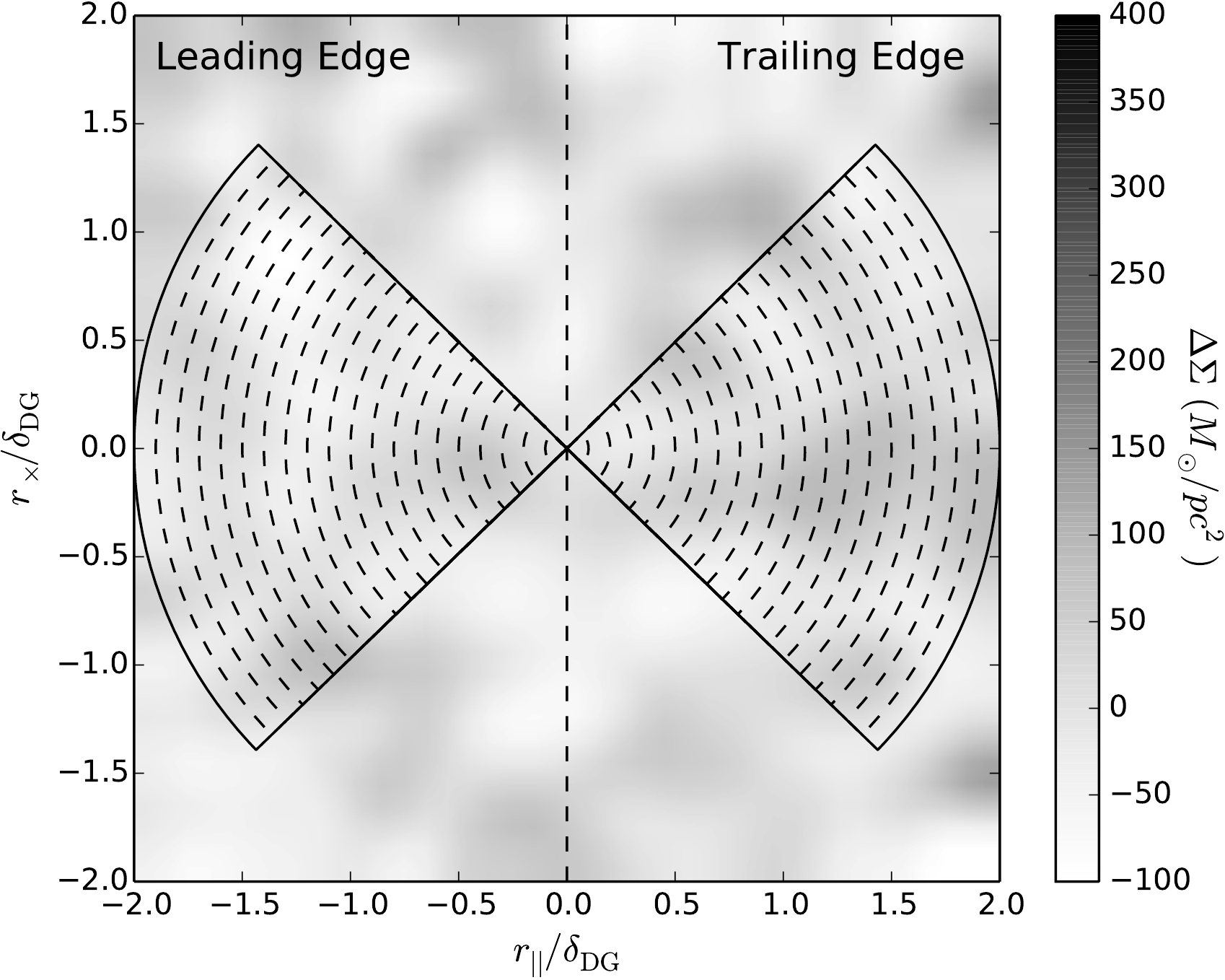} 
       	\caption{ The stacked results from 72 merging galaxy halos.
         		{\it Left :}  The stacked contribution for the projected mass density map, $\Sigma$, before the best-fit NFW halo is removed from each cluster field.
	               	{\it Right :} The stacked {\it residual} projected surface density map, $\Delta\Sigma$ after the best-fit NFW halo is removed from each cluster field. The cone shows the bins we use to project this map into the one dimensional profile, plus labels identifying lead and trail edges (the same as in Figure \ref{fig:draw_demo}).
	The distances are normalised to the separation of the gas and dark matter.}
	\label{fig:2d_maps}
\efigs
Figure \ref{fig:bias_constraints} shows the result of the Monte Carlo simulations. There are six panels with each one simulating a different source of uncertainty.
The top Figure of each panel shows the one dimensional density profile trailing (T) the dark matter, $\delta\Sigma_{\rm T}$. The simulated gas halo can be clearly seen at a distance of $1\dsg$ (by definition) from the origin. The small second simulated population of scattered particles can also be seen at $\sim0.2\dsg$, however in many of the simulations this is sub-systematic and cannot be resolved. The second Figure in each panel represents the difference between the trailing (T) and the leading (L) edge $\delta\Sigma_{\rm T}-\delta\Sigma_{\rm L}$. Each panel is in units of $M_\odot/$pc$^2$ and is simulating the following:
\begin{enumerate}
\item {\bf Top left } The effect of 2-dimensional statistical noise (x and y) on the estimate of the dark matter halo position. We find in order to resolve the second population of particles each halo must have a RMS of $<0.05\dsg$ ($\sim2.5\arcsec$). 
\item {\bf Middle left } The effect of a bias in the estimate of the typical size / scale radius of the dark matter halo. We find that although this affects the excess surface density profile along the tail, the difference between the tail and the lead is zero, and therefore the method is insensitive to this.
\item {\bf Bottom left} The effect of a bias in the estimate of the position of the gas halo. A large bias acts to smear out the gas halo bump and move the position of the substructure along the radial line. This will be important to handle in the case we attempt to interpret any excess in future experiments.
\item {\bf Top right} The effect of a bias in the measured position of the dark matter (positive towards the gas, negative away from it). 
Both the gas and any scattered DM will lag behind the un-scattered DM, and could systematically shift the estimated position for the parametric model in the trailing direction (since gravitational lensing probes all matter along the line of sight).  Subtracting off a symmetric model that lags behind the un-scattered DM would lead to the un-scattered DM contributing an excess surface density in the lead direction. Calculating the expected size of the shift in estimated position due to any gas or scattered DM is beyond the scope of this work. However here we find future experiments will require a bias of $\delta r_{||} < 0.1\dsg$ in order to make a detection.

\item {\bf  Middle right } The effect of a bias in the estimate of the normalisation of the dark matter halo. We find that the profile of the trail - lead excess surface density is left unbiased.
\item {\bf Bottom right} The effect of two-dimensional statistical error in the estimate gas position. This acts to smear out the position of the gas peak, and shift it slightly towards smaller $\dsg$. There is no observable effect on the second population of scattered particles.
\end{enumerate}
We conclude that the key error in this measurement is the precision with which we can estimate the position of each individual dark matter halo.
Given that we understand how the different systematics and statistical errors effect the method, we now apply it to data. 
 
 \section{The data}
We adopt the H15 sample of 30 galaxy clusters containing a total of 72 merging substructures.
These have been observed by the Hubble Space Telescope and Chandra X-ray Observatory.
For each substructure H15 measured the best-fitting mass profiles as described by the 6 NFW parameters, plus the X-ray positions and the flux weighted galaxy density distributions. 
We find for our sample that $\langle{\dsg}\rangle = 25 \pm 2\arcsec $, which means the free-form kappa map has a pixel size of $\delta_{\rm x} = \delta_{\rm y} = 0.1\dsg = 2.5\arcsec$ and a Gaussian smoothing kernel with a width of $7.5\arcsec$.
For more information on this dataset please see \cite{Harvey15}.

\section{Results}

\subsection{Two dimensional free-form surface density map}
We present the results of carrying out our new method on the 72 interacting cluster halos.
We test whether the potential interactions of dark matter produce an asymmetry in the dark matter profile. 
Figure \ref{fig:2d_maps} shows the excess surface density derived from the KS93 free-form method before (left-hand panel) and after (right-hand panel) we remove the best-fit NFW model using {\sc Lenstool}. 
We also show in the right hand panel the binning we use to project this two-dimensional map into a one-dimensional profile in the next section plus labels identifying the leading and trailing edge of the merging clusters.
We clearly see in the left hand panel the mean mass profile from the ensemble of clusters before the best fit model is removed, and the right hand panel that appears to be consistent with noise.

\subsection{One dimensional residual surface density profile}
In order to further test whether it is consistent with noise, we project this two dimensional free-form residual surface density map (right hand panel of Figure \ref{fig:2d_maps}) into one dimension along the axis of collision, examining the profile in vertical bins along the $r_{||}$ axis, taking care only to bin up to the $45^\circ$ axis dividing $r_{||}$ and $r_\times$.
The black points in the top panel of Figure \ref{fig:profiles_compare} shows the results from the stacked data for $\delta\Sigma_{\rm T}$, the projected residual surface density profile. The black points in bottom panel of Figure \ref{fig:profiles_compare} gives the difference between the trailing and leading surface density, with the subscripts `T' and `L' referring to profile Trailing the halo and Leading the halo. 
We find that there is an excess surface density around $r=0.5\dsg$  at the $\sim2\sigma$ level and $r=1.25\dsg$ at the $\sim4\sigma$ level. 
The significance of the first bump disappears when comparing the trail and leading edge, however the second bump becomes more prominent.
This excess is sufficiently distant to be consistent with baryonic gas that has been stripped during the collisions, however its offset from $\dsg=1$ is curious. 

To understand the shape of this result we use our simulations from our systematics test in section \ref{sec:errors} except add an additional free parameter which is gas mass fraction. This results in seven free parameters; the statistical and systematic error in the dark matter position, the systematic bias in the estimate of the radius and amplitude size of the dark matter halo, the statistical and systematic error in the estimate of the position of the gas peak, the gas mass fraction and the fraction of scattered particles that are in a second population trailing the dark matter. We run a Monte Carlo Markov Chain (MCMC) with Metropolis Hastings sampling, simultaneously fitting both the trail data and the trail - lead data (i.e. both the panels in Figure \ref{fig:profiles_compare}). We sample to find the best fit noise parameters. 
The grey regions in Figure \ref{fig:profiles_compare} show the best fit model with the associated one-sigma error. 
Even with simple Gaussian assumptions we can reproduce the shape of the data well.
We find that that the data best fits a model with 
statistical scatter in the position of dark matter of $0.15\pm0.07\dsg$, 
systematic offset in the position of dark matter of $0.0002\pm0.0006\dsg$,
a bias in the amplitude of the dark matter of  $-9\pm10\%$, 
a systematic error in the radius of the dark matter halos of  $-32\pm8\%$, 
and a statistical and systematic error in the position of the gas peak of $0.08\pm0.2\dsg$ and $0.23\pm0.03\dsg$ respectively. 
We estimate the trailing gas mass fraction $M_{\rm gas} / M_{ \rm dm} = 0.13\pm0.035$.
Finally we find no evidence for a second population of particles, with the fraction at $0.03\pm0.05$.

The sensitivity of our data and precision of the dark matter mapping means that we are not sensitive to $\sigma_{\rm DM} \le 1$cm$^2$/g. In order to be so we require our statistical error in the position of the dark matter halo to be three times smaller. However, with the error bars deduced in Figure \ref{fig:profiles_compare} we can predict the power of future experiments.

\fig
       \includegraphics[width=0.5\textwidth]{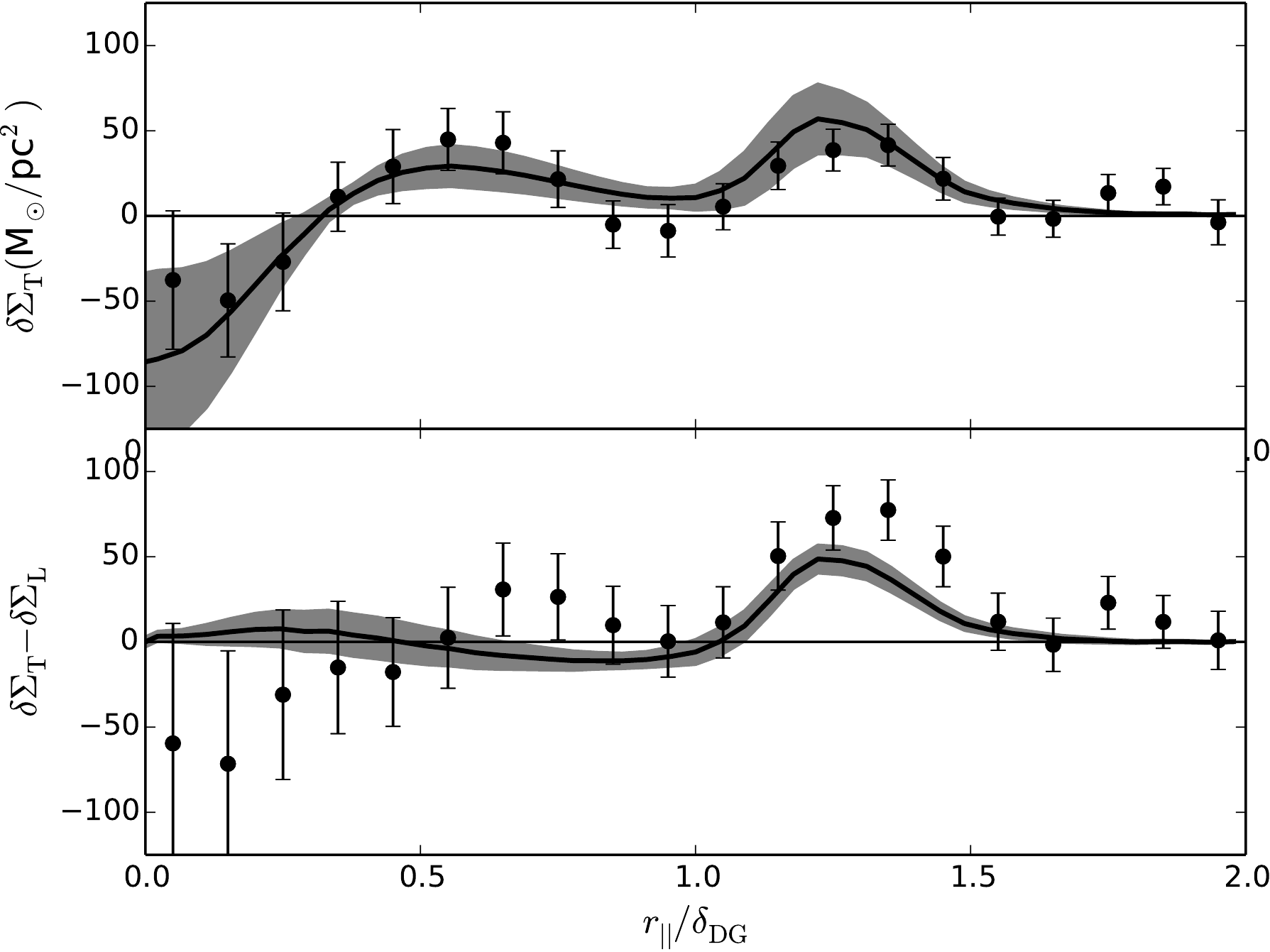}
       \caption{ 	\label{fig:profiles_compare}The projected profile of the stacked excess surface density map in Figure \ref{fig:2d_maps}. The black points are the results from the stacked data. The black solid line represents the best fit error-model of all the associated uncertainties with the grey region showing the one-sigma error in this model.}
\efig
 
   \fig
      \vspace{-0.65cm}
                     \includegraphics[width=0.55\textwidth]{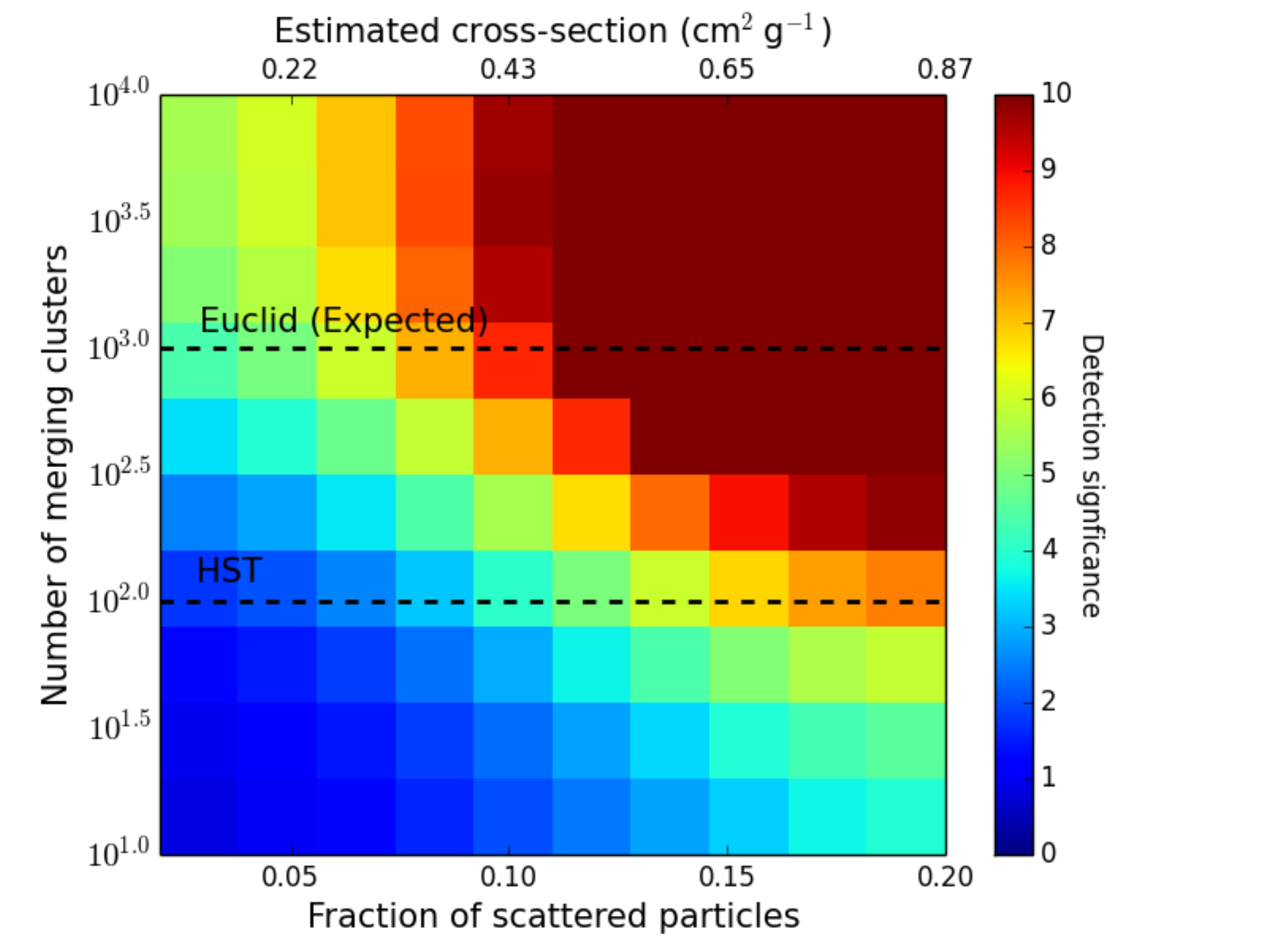} 
       	\caption{ The expected detection (statistical) significance of a second population of scattered particles for a given scattering fraction  and sample size (assuming we can estimate the position of a single dark matter halo to $<0.05\dsg$). We highlight the estimated expected sample sizes from the Hubble Space Telescope and the forthcoming Euclid mission.  }
	\label{fig:constraints}
\efig

\section{Future prospects}

To estimate the statistical power of this technique for future surveys we generate a second population of particles as before, except with varying scattering fractions. 
Assuming a linearity between number of scattered particles and cross-section and that all particles are scattered in to the tail of the halo, we can estimate the probability of a detection for a given cluster sample size. For each scattering fraction and sample size we simulate an observation. We then fit a model to the simulated data and through an MCMC determine the detection significance. Figure \ref{fig:constraints} shows the results. The bottom x-axis shows the simulated scattering fraction, and the top x-axis shows the respective estimated cross-section. The black dotted lines show the sample size expected by the end of the Hubble Space Telescope lifetime, and the upcoming Euclid mission \citet{EUCLID}. We find that with HST we could be able to detection scattering fractions of $<10\%$ at the $3\sigma$ level. However in order to reach the precision required in the dark matter models we will require the sample to have high quality redshift estimations and cluster member identification.

\section{Conclusions}
We test the hypothesis that dark matter interacts through non-gravitational forces and is therefore scattered towards the rear of a cluster during collisions creating asymmetry in the distribution of dark matter along the merger axis.
Using the sample of 30 merging galaxy clusters observed by the Hubble Space Telescope and Chandra X-ray Observatory used in \cite{Harvey15}, we fit and subtract the best-fitting symmetric NFW halo.
We measure the residual distribution of mass not accounted for by the NFW fit.
We project the surface density along the axis of collision and we test for residual mass leading or trailing each cluster, and perform the same measurement along the  perpendicular axis as a null test.
We detect at $4\sigma$ significance, a mass peak at $1.23\dsg$, which we attribute to the gas mass in the cluster, unaccounted for by the lensing model.
We measure the gas fraction at $M_{\rm gas} / M_{ \rm dm} = 0.13\pm0.035$.
Through numerical simulations of Gaussian halos we find that the one dimensional excess profile can be well fit by a simple model of the noise, and find that we can reproduce the shift in the gas mass if we have  a statistical and systematic error in the position of the gas peak of $0.08\pm0.2\dsg$ and $0.23\pm0.03\dsg$ respectively. This model finds no evidence for any second population of scattered particles with an scattering fraction estimate of $f=0.03\pm0.05$.
Furthermore we find that the limiting factor in this method is the statistical precision to which we can estimate the position of the dark matter halo. 
Future studies will require positional precision of $<0.05\dsg$ ($\sim<2.5\arcsec$), however should we meet this requirement, current samples from the Hubble Space Telescope could have the potential to detect scattering fractions of $<10\%$ ($\sdm\simlt0.4$\cmg) at the $>3\sigma$ level.
In order to interpret any excess over the noise model we will require hydrodynamical simulations with self-interacting dark matter simulations.

We present here an interesting method that shows promise for {\it current} and future surveys. In a bid to measure and confirm detections of self-interacting dark matter, this method provides an independent test for scattering dark matter that will be required in the event of any detection.

\section*{Acknowledgments}
DH is supported by the Swiss National Science Foundation (SNSF). JPK acknowledges support from the ERC advanced grant LIDA and from CNRS.
AR was supported by the UK Science and Technology Facilities Council grant numbers ST/K501979/1 and ST/L00075X/1. RM was supported by the Royal Society.
\bibliographystyle{mn2e}
\bibliography{bibliography}

\bsp

\label{lastpage}

\end{document}